\documentclass[prb, twocolumn]{revtex4}
\usepackage{graphicx}
\begin{document}
\title{\flushleft{Dynamics of propagating turbulent pipe flow
    structures. Part II:  Relaminarization}}
\date{\today}
\author{A. Duggleby}
\author{K.S. Ball}
\author{M.R. Paul}
\affiliation{Department of Mechanical Engineering, Virginia Polytechnic Institute and State University Blacksburg, Virginia  24061}
\email{duggleby@vt.edu}

\begin{abstract}
\noindent The dynamical behavior of propagating structures, determined
 from a Karhunen-Lo\`{e}ve decomposition, in turbulent pipe flow
 undergoing reverse transition to laminar flow is investigated.
The turbulent flow data is generated by a direct numerical simulation started
at a fully turbulent Reynolds number of $\mathrm{Re}_\tau=150$, which is
slowly decreased until
$Re_\tau=95$.  At this
low Reynolds number the high frequency modes decay first, leaving only
the decaying streamwise vortices.  The flow undergoes a chugging phenomena, where it
begins to relaminarize and the mean velocity increases.  The
remaining propagating modes then destabilize the
streamwise vortices, rebuild the energy spectra, and eventually the flow regains
its turbulent state.  Our results capture three chugging cycles before 
the flow completely relaminarizes.  The high frequency
modes present in the outer layer decay first, establishing the
importance of the outer region in the self-sustaining mechanism of
wall bound turbulence.
\end{abstract}

\maketitle

\section{INTRODUCTION}

In part I the effect of drag reduction on propagating turbulent flow
structures by spanwise wall oscillation
was studied. \cite{duggleby_drPipe}  A second instance where drag reduction is seen is in a
relaminarizing flow.  As the turbulence dies, so does the Reynolds
stress generation, and thus, for a constant pressure gradient driven
flow, the flow rate increases.  As we will show, the flow does not
immediately relaminarize, but instead goes through a series of
chugging motions.  In these chugging motions, the flow loses its
turbulent inertial range, losing the high frequencies first.  Before the
flow has completely relaminarized, certain key propagating waves
interact with the decaying streamwise vortices, recreating the
cascading energy scales that populate the
inertial subrange.  In this part, we examine the dynamics found in
relaminarization to understand how a flow remains turbulent to better
elucidate the mechanism behind the self-sustaining nature of
turbulence.

Previous work has focused either on
relaminarization from favorable pressure gradients, \cite{talamelli,
  fernholz98a, fernholz98b} strong accelerations \cite{greenblatt}, or
 examining relaminarization as a
testbed to understand the decay rate of structures. \cite{peixinho}
Outside the field of wall turbulence, there has been related work in
studying relaminarization and 
bifurcations in spherical Couette flow. \cite{nakabayashi}  The
only work found that relates to drag reduction is the examination of
linear feedback control in a turbulent channel flow that achieved
total relaminarization. \cite{hogberg}  Nevertheless, it is of interest
to examine the field of transition where recent work has added to our
knowledge of pipe flow structures and their interactions.

Both Kerswell \cite{kerswell} and Faisst and Eckhardt \cite{faisst} have
found traveling wave solutions to the Navier-Stokes equations through
continuation methods.  They identified structures for rotationally
symmetric solutions, which is confirmed here for
$\mathrm{Re}_\tau=95$ and in previous work \cite{duggleby_JOT} for $\mathrm{Re}_\tau=150$ through a
Karhunen-Lo\`{e}ve (KL) decomposition of a direct numerical simulation of
turbulent pipe flow.  Moreover, Kerswell found that the threefold
rotation, or azimuthal wavenumber $n=3$ in our notation, is the
largest contributor near the critical bifurcation point associated
with the laminar to turbulent transition.  This also corresponds to
our findings, as we will show that when observing the chugging phenomena, the $n=3$ traveling wave is often
the most energetic at the point which the flow reasserts itself as
turbulent.

Also of note is the minimal channel work
by Webber, Handler, and Sirovich; we apply their results to further
the understanding of this chugging phenomena. \cite{webber2}  They indicate that
the nonlinear terms in the Navier-Stokes equations lead to triad
interactions of the KL modes which is responsible for the transfer of
energy between modes.  This occurs whenever wavenumbers of three modes $(m,n,q), (m',
n', q')$, and $(m'', n'', q'')$ sum to zero, shown in equations
\ref{triadN} and \ref{triadM} below,

\begin{eqnarray}
\label{triadN}
& n+n'+n''=0 \\
\label{triadM}
& m+m'+m''=0 
\end{eqnarray}
where n is the azimuthal wavenumber and m is the streamwise wavenumber
obtained from the Fourier representation of the flow.

\section{NUMERICAL METHOD}
Details on the numerical method for generating the direct numerical
simulation (DNS) flow fields and
on the Karhunen-Lo\`{e}ve method can be found in  part I of this paper.
\cite{duggleby_drPipe}

\begin{figure}
\centering
\includegraphics[width=3.25 in]{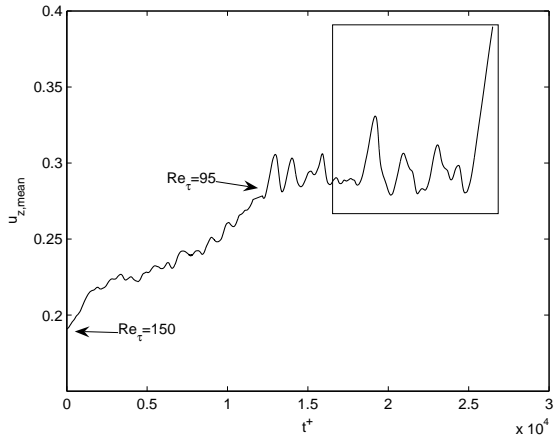}
\caption{Time history of the mean flow rate relaminarization 
  showing the initialization from $\mathrm{Re}_\tau=150$ to
  $\mathrm{Re}_\tau=95$.  The window shows where data was collected.}
\label{meanhistFull}
\end{figure}

\begin{figure}
\centering
\includegraphics[width=3.25 in]{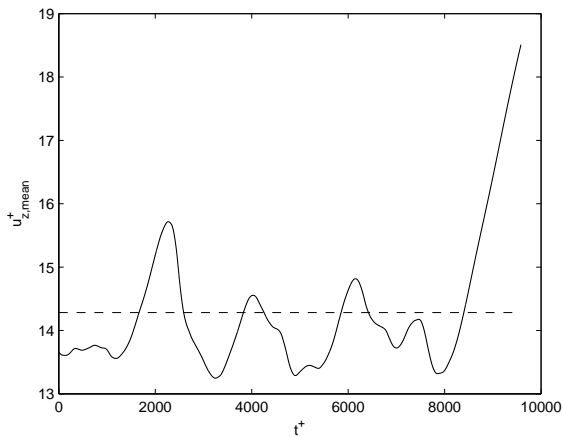}
\caption{Time history of the mean flow rate relaminarization for
  $\mathrm{Re}_\tau=95$.  Three chugging cycles are seen near
  $t^+=2000,4000$ and $6000$ with the final relaminarization starting
  near $t^+=8000$.}
\label{meanhist}
\end{figure}

Over a time of 12,000 $t^+$ the Reynolds number was slowly reduced
from $\mathrm{Re}_\tau=150$ to a value of $\mathrm{Re}_\tau=95$, and
the DNS continued for another $5000 t^+$
to eliminate any transient effects, as seen in Figure \ref{meanhistFull}.  Data was then collected for
$10,000 t^+$, which included three distinct chugs before the flow completely relaminarized, as seen in Figure
\ref{meanhist}.

The grid resolution was kept the same as for the original
$\mathrm{Re}_\tau=150$ case.  Thus the grid is effectively further refined
to $\Delta r^+\approx 0.49$ and $(R \Delta \theta)^+ \approx 3.1$ near the
wall and $\Delta^+ \approx 2.0$ near the centerline with
a constant streamwise resolution of $\Delta z^+=4.0$, where $r$,
$\theta$, and $z$ is the radial, azimuthal, and streamwise direction,
respectively, and $R$ is the radius of the pipe.

\section{RESULTS}

\begin{figure}
\centering
\includegraphics[width=3.5 in]{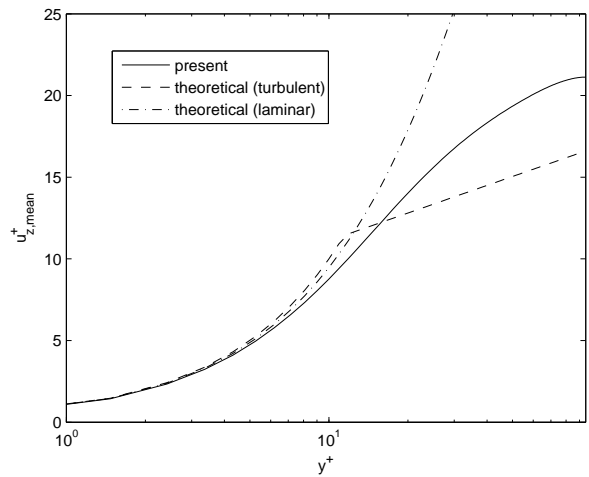}
\caption{Mean flow profile versus $y^+=(1-r)\mathrm{Re}_\tau$
  including the theoretical turbulent profile (dashed) with the
  sublayer ($u^+=y^+$), the log layer ($u^+=\log (y^+)/0.41+5.5$),
  and the analytical parabolic laminar solution (dash-dot). The
  mean flow profile follows the law of the wall, yet 
  deviates drastically from the log layer as expected.}
\label{meanLog}
\end{figure}

The profile of the mean flow with respect to wall units is seen in
Figure \ref{meanLog}.  At this low Reynolds number, the profile does
not conform to the log layer, yet near the wall it still exhibits a linear
sublayer.  Also, the flow does not conform to the laminar parabolic
profile, indicating that it is indeed turbulent.

\begin{figure}
\centering
\includegraphics[width=3.5 in]{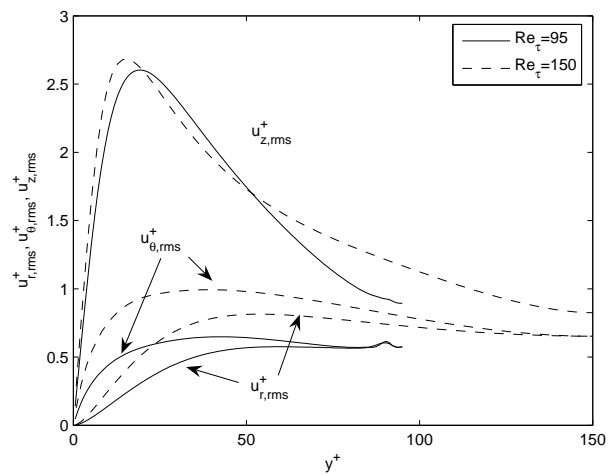}
\caption{Root-mean-squared velocity profiles for $\mathrm{Re}_\tau=95$ (solid)
  and $\mathrm{Re}_\tau=150$ (dashed) versus wall units $y^+$.  The
  inflections near $y^+=90$ are effects of the laminar chugging, because
  the rms velocities are averaged over all time steps. }
\label{rms}
\end{figure}

The root-mean-square (rms) velocity profiles also show a turbulent trend as
seen in Figure \ref{rms},
although in comparison to the $\mathrm{Re}_\tau=150$  flow, the radial and
azimuthal fluctuations are about half as strong, and the peak streamwise
rms velocity is shifted further away from the wall.  Also
noteworthy is the strong fluctuations near the center of the pipe at $y^+
\approx 90$.  This is the first indication that the dynamics near the
center of the pipe differ from what is expected for the fully turbulent case.

\begin{figure}
\centering
\includegraphics[width=3 in]{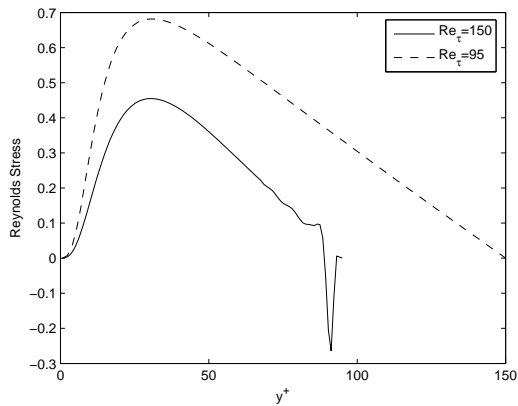}
\caption{Reynolds stress profiles for $\mathrm{Re}_\tau=95$ (solid)
  and $\mathrm{Re}_\tau=150$ (dashed) versus wall units $y^+$.  As in
  the rms velocity profiles, the inflections near $y^+=90$ are effects of the
  laminar chugging, showing that relaminarization begins at the center
  of the pipe.}
\label{reynolds}
\end{figure}

The Reynolds stress profile also differs from the
$\mathrm{Re}_\tau=150$ case and is shown in Figure \ref{reynolds}.
The Reynolds stress for the 
$\mathrm{Re}_\tau=95$ case has roughly  half the magnitude
throughout the radial profile, although the peak Reynolds stress is
also found at the same location of $y^+=31$.  The largest
deviation from the expected profile occurs near the centerline beyond $y^+ \approx 60$, where the
Reynolds stress begins to fluctuate.  In particular, between $y^+=88$
and $92$, the Reynolds stress is negative, which is physically
interpreted as turbulence damping.  Now, in addition to the
rms deviation near the centerline, this Reynolds stress fluctuation
indicates that the relaminarization process begins at the center of
the pipe and goes towards the wall.  

\begin{figure}
\centering
\includegraphics[width=3 in]{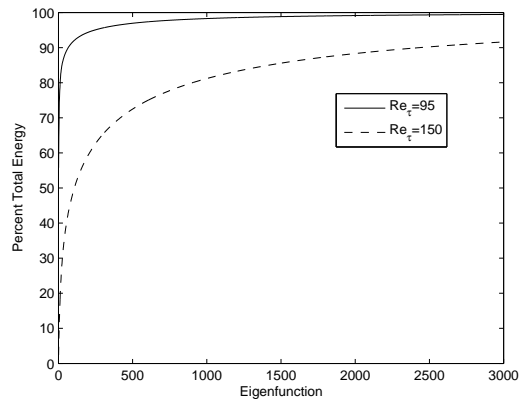}
\caption{Comparison of the running total energy retained in the KL
  expansion for $\mathrm{Re}_\tau=95$ (solid)
  and $\mathrm{Re}_\tau=150$ (dashed).  The 90\% crossover point contains
  2453 and 66 modes respectively, showing a drastic reduction in the 
  turbulent attractor.}
\label{KLdim}
\end{figure}

Turning from statistics to the KL decomposition, we find that the chaotic attractor is reduced in size, as expected with a reduction
in $\mathrm{Re}_\tau$, from $D_{KL}=2453$ to $D_{KL}=66$, shown in
Figure \ref{KLdim}.  This dimension is similar to that found in Part I where the oscillated pipe was barely turbulent with a
dimension of $D_{KL}=102$, and any stronger oscillation would have
resulted in relaminarization.

In observing the energy content of the modes in Table \ref{eigenList},
and the most increased and decreased in Table \ref{energy_change}, we find that the
shear modes increase drastically, which is a result of the chugging
motion and large mean flow rate fluctuations.  Also of note is the
increase in strength of the  $n=3,4$ and $5$  streamwise vortices $(m=0)$ and
their associated wall traveling waves $(m=1)$.  The increase in the
$(0,1,1)$ and $(1,0,1)$ modes, since they are not found in the
work by Kerswell \cite{kerswell} and Faisst and Eckhardt,
\cite{faisst} could be the catalysts of energy triads in
equations \ref{triadN} and \ref{triadM} between the $n=3,4$ and $5$
rolls and the wall traveling waves.  For example, the $(1,3,1)$ and $(1,4,1)$
waves interact through the $(0,1,1)$ catalyst, and the $(1,3,1)$ wave and
the $(0,3,1)$ roll interact through the $(1,0,1)$ catalyst.  The modes
that decreased the most in energy are more of a result of the
lower Reynolds number, as the high modes $n=8,9$ and $10$ in the
$\mathrm{Re}_\tau=150$ case were an important basis for streamwise vortices that
were $94-117$ wall units apart.  However, at $\mathrm{Re}_\tau=95$,
these represent spacings of $60-75$ wall units, which is too small,
thus explaining their drop in energy.

\begin{table}
\caption{Comparison of first 25 eigenvalues.  $m$ is the streamwise
  wavenumber, $n$ is the spanwise wavenumber, and $q$ is the eigenvalue
  quantum number.}
\label{eigenList}
\centering
\begin{tabular}{lcccclcccccc}
\hline
\hline
& \multicolumn{5}{c}{$\mathrm{Re}_\tau=150$}  & &
\multicolumn{5}{c}{$\mathrm{Re}_\tau=95$}  \\
Index & $m$ & $n$ & $q$ &  Energy & \% Total & & $m$ & $n$ & $q$ & Energy & \%
Total \\
\hline
1 & 0 & 6 & 1 & 1.61 & 2.42\%  & & 0 & 0 & 1 & 114 & 55.76\% \\
2 & 0 & 5 & 1 & 1.48 & 2.22\%  & & 0 & 3 & 1 & 9.63 & 4.72\%\\
3 & 0 & 3 & 1 & 1.45 & 2.17\%  & & 0 & 4 & 1 & 8.39 & 4.11\%\\
4 & 0 & 4 & 1 & 1.29 & 1.93\%  & & 0 & 1 & 1 & 7.79 & 3.82\%\\
5 & 0 & 2 & 1 & 1.26 & 1.88\%  & & 0 & 2 & 1 & 6.30 & 3.09\%\\
6 & 1 & 5 & 1 & 0.936 & 1.40\% & & 0 & 0 & 2 & 5.98 & 2.93\%\\
7 & 1 & 6 & 1 & 0.917 & 1.37\% & & 0 & 5 & 1 & 2.79 & 1.37\%\\
8 & 1 & 3 & 1 & 0.902 & 1.35\% & & 1 & 4 & 1 & 2.18 & 1.07\%\\
9 & 1 & 4 & 1 & 0.822 & 1.23\% & & 1 & 3 & 1 & 1.85 & 0.90\%\\
10 & 0 & 1 & 1 & 0.805 & 1.20\% & & 0 & 0 & 3 & 1.83 & 0.90\% \\
11 & 1 & 7 & 1 & 0.763 & 1.14\% & & 0 & 6 & 1 & 1.56 & 0.77\%\\
12 & 1 & 2 & 1 & 0.683 & 1.02\% & & 1 & 5 & 1 & 1.45 & 0.71\%\\
13 & 0 & 7 & 1 & 0.646 & 0.97\% & & 1 & 2 & 1 & 1.42 & 0.70\%\\
14 & 2 & 4 & 1 & 0.618 & 0.92\% & & 0 & 1 & 2 & 1.34 & 0.66\%\\
15 & 0 & 8 & 1 & 0.601 & 0.90\% & & 1 & 0 & 1 & 1.30 & 0.64\%\\
16 & 2 & 5 & 1 & 0.580 & 0.87\% & & 1 & 6 & 1 & 0.884 & 0.43\%\\
17 & 1 & 1 & 1 & 0.567 & 0.85\% & & 1 & 1 & 1 & 0.815 & 0.40\%\\
18 & 2 & 7 & 1 & 0.524 & 0.78\% & & 0 & 7 & 1 & 0.727 & 0.36\%\\
19 & 1 & 8 & 1 & 0.483 & 0.72\% & & 2 & 4 & 1 & 0.716 & 0.35\%\\
20 & 2 & 6 & 1 & 0.476 & 0.71\% & & 2 & 3 & 1 & 0.665 & 0.33\%\\
21 & 2 & 3 & 1 & 0.454 & 0.68\% & & 2 & 5 & 1 & 0.589 & 0.29\%\\
22 & 2 & 2 & 1 & 0.421 & 0.63\% & & 2 & 2 & 1 & 0.544 & 0.27\%\\
23 & 2 & 8 & 1 & 0.375 & 0.56\% & & 0 & 2 & 2 & 0.521 & 0.26\%\\
24 & 1 & 9 & 1 & 0.358 & 0.54\% & & 2 & 6 & 1 & 0.487 & 0.24\%\\
25 & 3 & 4 & 1 & 0.354 & 0.53\% & & 1 & 1 & 2 & 0.479 & 0.23\%\\
\hline
\hline
\end{tabular}
\label{top25}
\end{table}
\begin{table}
\caption{Ranking of eigenfunctions by energy change between the
  $\mathrm{Re}_\tau=150$ and $\mathrm{Re}_\tau=95$ cases.  $m$ is the streamwise
  wavenumber, $n$ is the spanwise wavenumber, and $q$ is the eigenvalue
  quantum number.  The shear modes $(0,0,q)$ are among the most
  increased, as are the  $n=3,4,$and $5$ rolls $(m=0)$ and wall modes $(m=1)$. }
\label{energy_change}
\centering
\begin{tabular}{llcccllccc}
\toprule
 &  \multicolumn{4}{c}{Increase} & & \multicolumn{4}{c}{Decrease} \\
Rank & $\Delta \lambda_{\mathbf{k}}$ & $m$ & $n$ & $q$ & & $\Delta \lambda_{\mathbf{k}}$ & $m$ & $n$ & $q$ \\
\colrule
1 & 113.4 & 0 & 0 & 1 & &-0.399 & 1 & 7 & 1 \\
2 & 8.18 & 0 & 3 & 1 & &-0.300 & 0 & 8 & 1 \\
3 & 7.10 & 0 & 4 & 1 & &-0.241 & 1 & 8 & 1 \\
4 & 6.99 & 0 & 1 & 1 & &-0.236 & 2 & 7 & 1 \\
5 & 5.81 & 0 & 0 & 2 & &-0.229 & 1 & 9 & 1 \\
6 & 5.04 & 0 & 2 & 1 & &-0.225 & 2 & 8 & 1 \\
7 & 1.71 & 0 & 0 & 3 & &-0.186 & 0 & 9 & 1 \\
8 & 1.36 & 1 & 4 & 1 & &-0.175 & 3 & 9 & 1 \\
9 & 1.30 & 0 & 5 & 1 & &-0.173 & 3 & 8 & 1 \\
10 & 1.12 & 0 & 1 & 2 & & -0.170 & 2 & 9 & 1 \\
11 & 1.09 & 1 & 0 & 1 & & -0.165 & 1 & 10 & 1 \\
12 & 0.94 & 1 & 3 & 1 & & -0.149 & 0 & 10 & 1 \\
\botrule
\end{tabular}
\end{table}

\begin{figure}[tb]
\includegraphics[width=3 in]{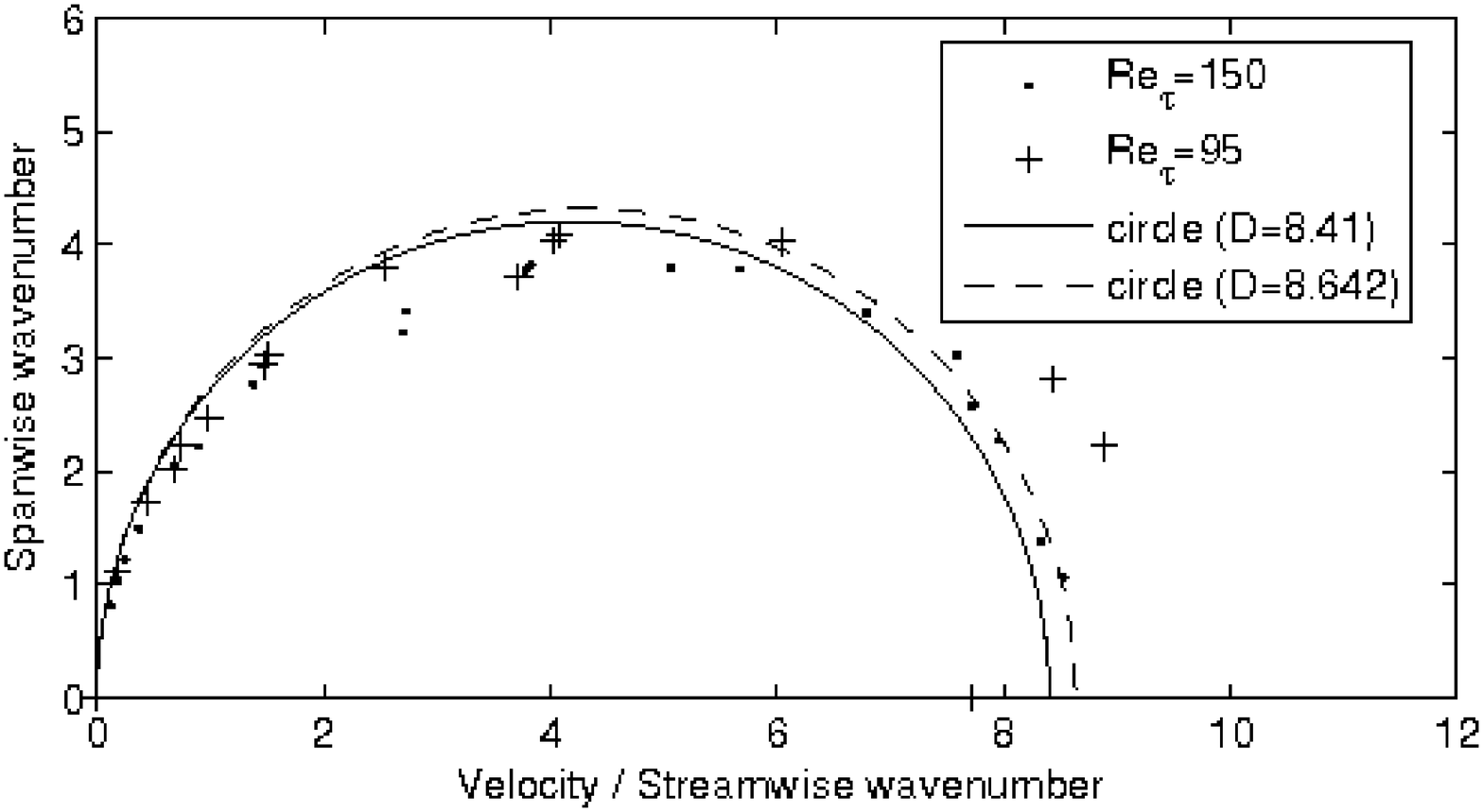}
\caption{Comparison of normal speed locus for the $\mathrm{Re}_\tau=95$ ($\cdot$) and
 $\mathrm{Re}_\tau=150$ ($+$) cases.  The solid lines represent a circle of
diameter 8.41 and 8.64 respectively that intersect at the origin.}
\label{locus}
\end{figure}

Looking at the normal speed locus in Figure \ref{locus}, the modes for a wave packet similar
to that of
$\mathrm{Re}_\tau=150$, but with a slightly
faster advection speed of $8.64$ versus $8.41$.  This shows a Reynolds number
dependence on the advection speed, as expected, since the advection speed
was shown to scale with the mean flow rate when the flow rate was
increased in part I with spanwise wall oscillation.

\begin{figure}[htb]
\begin{tabular}{cc}
\begin{minipage}{1.75 in}
\includegraphics[width=1.75 in]{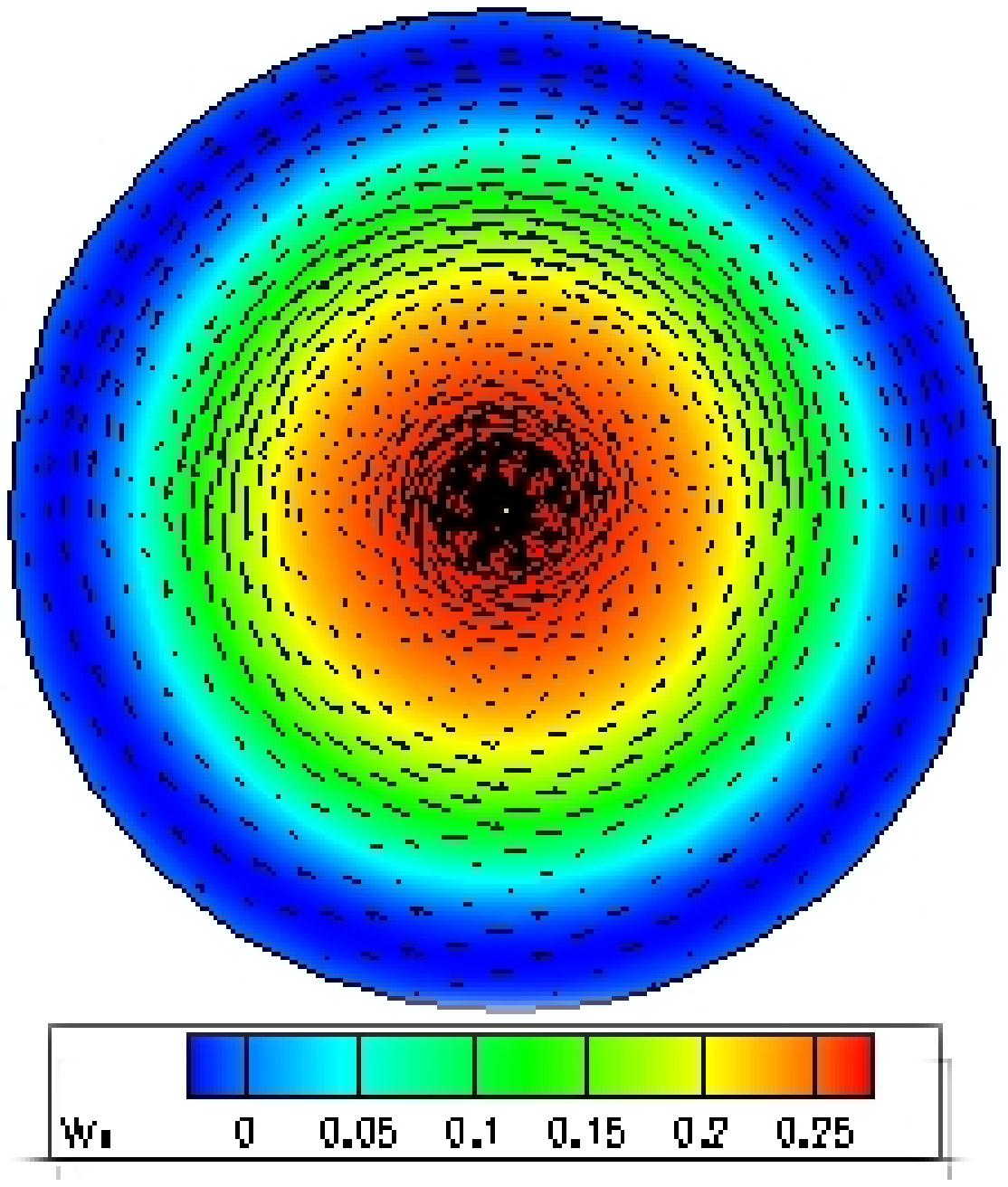}
\end{minipage}
&
\begin{minipage}{1.75 in}
\includegraphics[width=1.75 in]{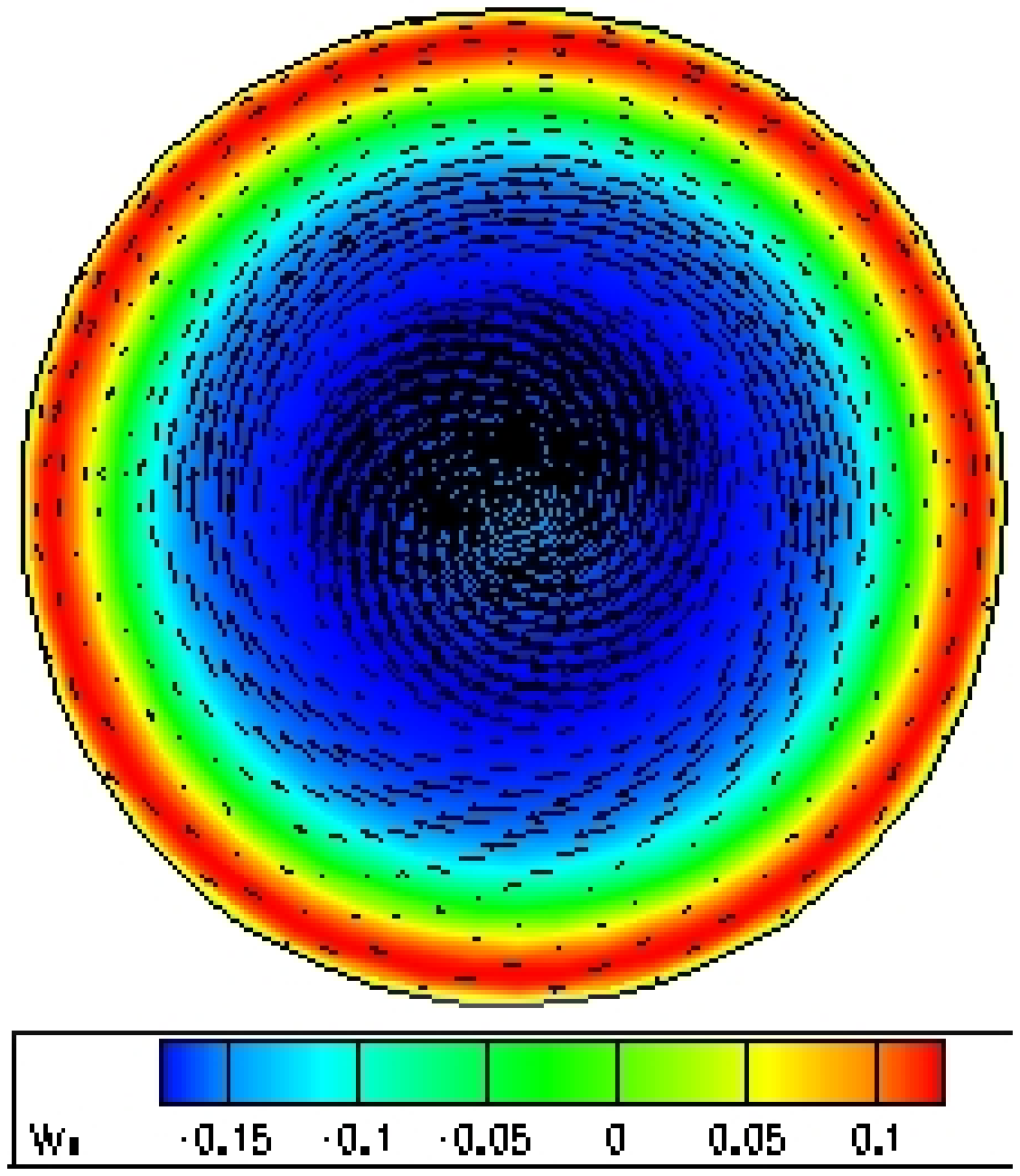}
\end{minipage} \\
\end{tabular}
\caption{The (0,0,1) shear mode with contours of streamwise velocity and
  vectors of cross-stream velocities. Left: (a) $\mathrm{Re}_\tau=95$.
  Right: (b) $\mathrm{Re}_\tau=150$.}
\label{001}
\end{figure}

\begin{figure}[htb]
\begin{tabular}{cc}
\begin{minipage}{1.75 in}
\includegraphics[width=1.75 in]{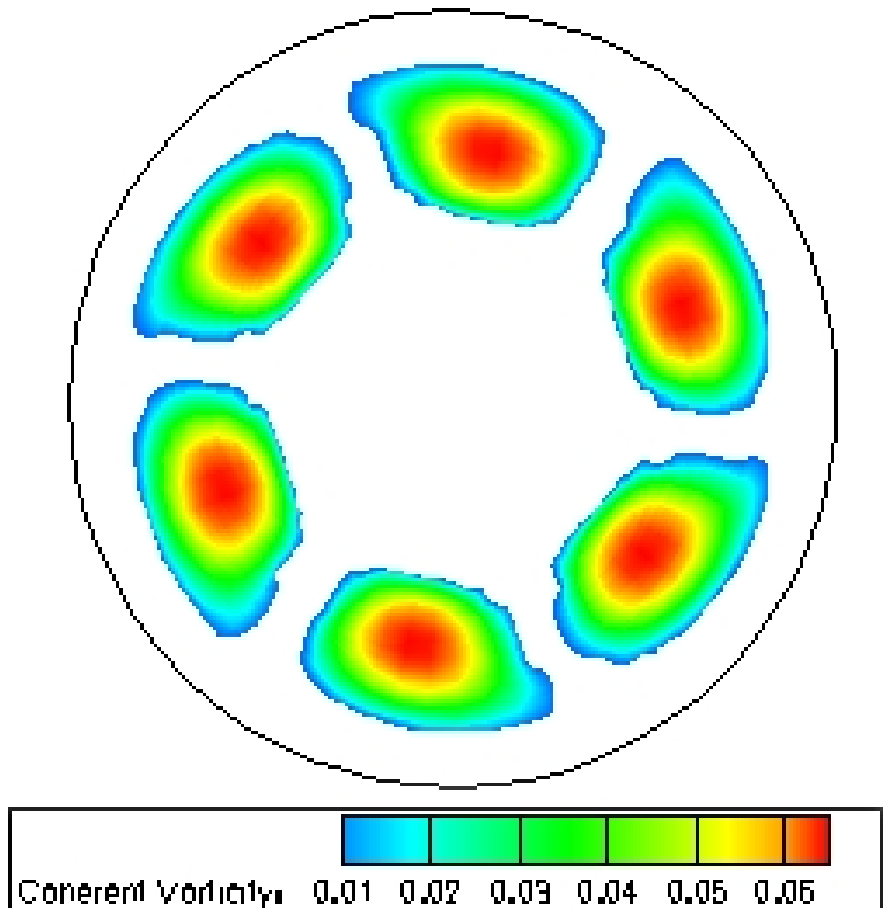}
\end{minipage}
&
\begin{minipage}{1.75 in}
\includegraphics[width=1.75 in]{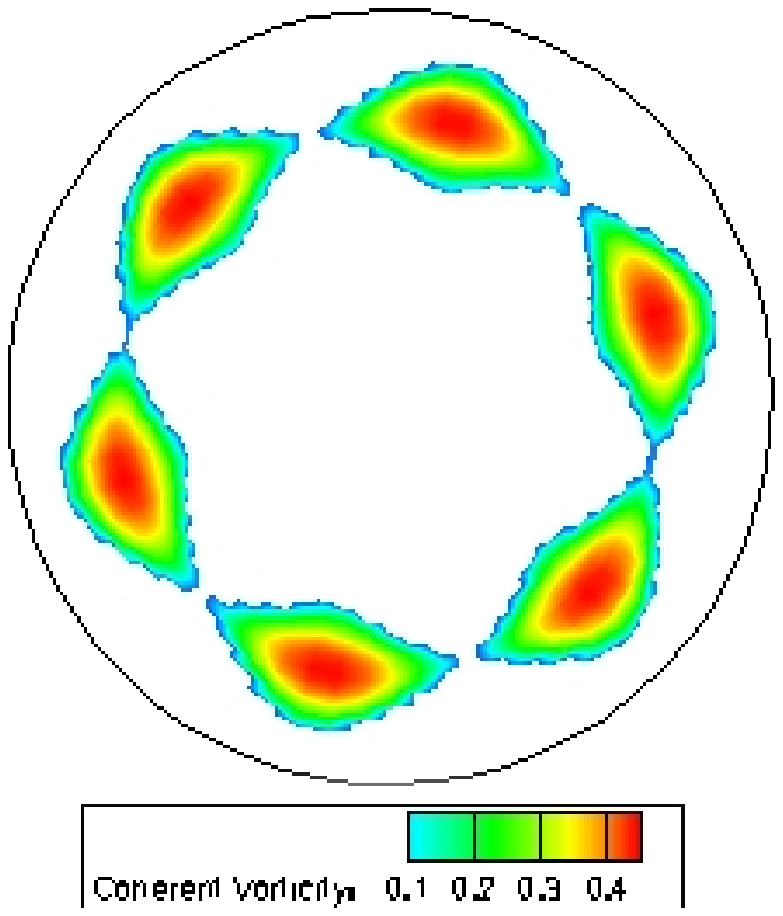}
\end{minipage}\\
\end{tabular}
\caption{The (0,3,1) roll mode with contours of coherent vorticity. Left: (a) $\mathrm{Re}_\tau=95$.
  Right: (b) $\mathrm{Re}_\tau=150$.}
\label{031}
\end{figure}

\begin{figure}[htb]
\begin{tabular}{cc}
\begin{minipage}{1.75 in}
\includegraphics[width=1.75 in]{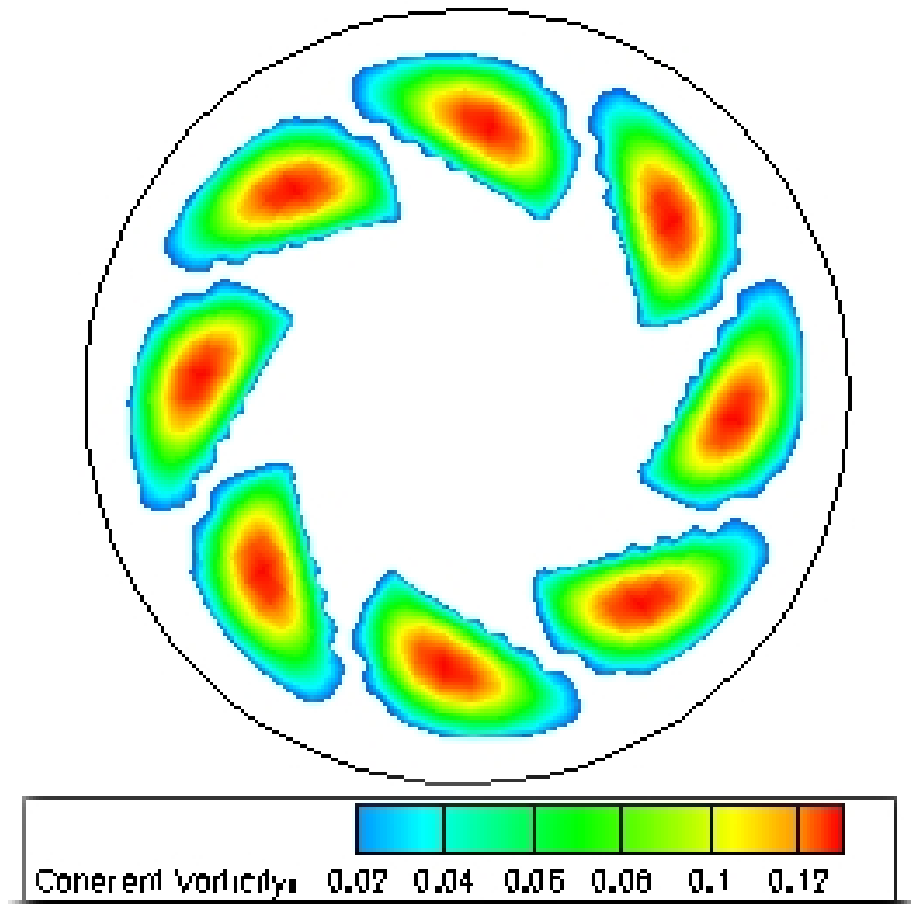}
\end{minipage}
&
\begin{minipage}{1.75 in}
\includegraphics[width=1.75 in]{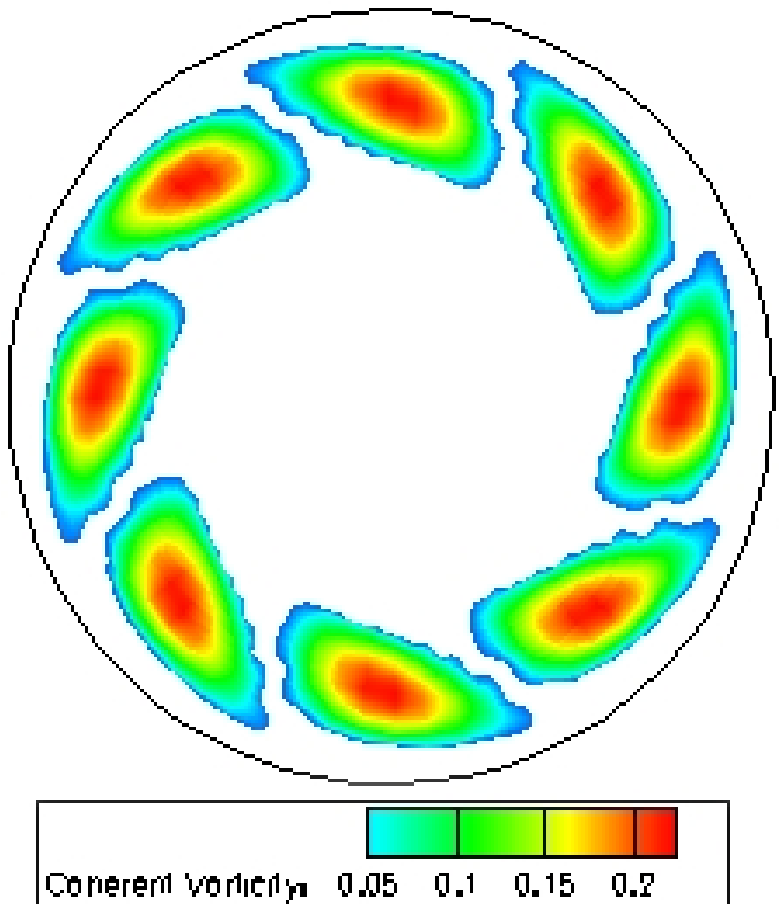}
\end{minipage} \\
\end{tabular}
\caption{The (1,4,1) wall mode with contours of coherent vorticity. Left: (a) $\mathrm{Re}_\tau=95$.
  Right: (b) $\mathrm{Re}_\tau=150$.}
\label{141}
\end{figure}

\begin{figure}[htb]
\begin{tabular}{cc}
\begin{minipage}{1.75 in}
\includegraphics[width=1.75 in]{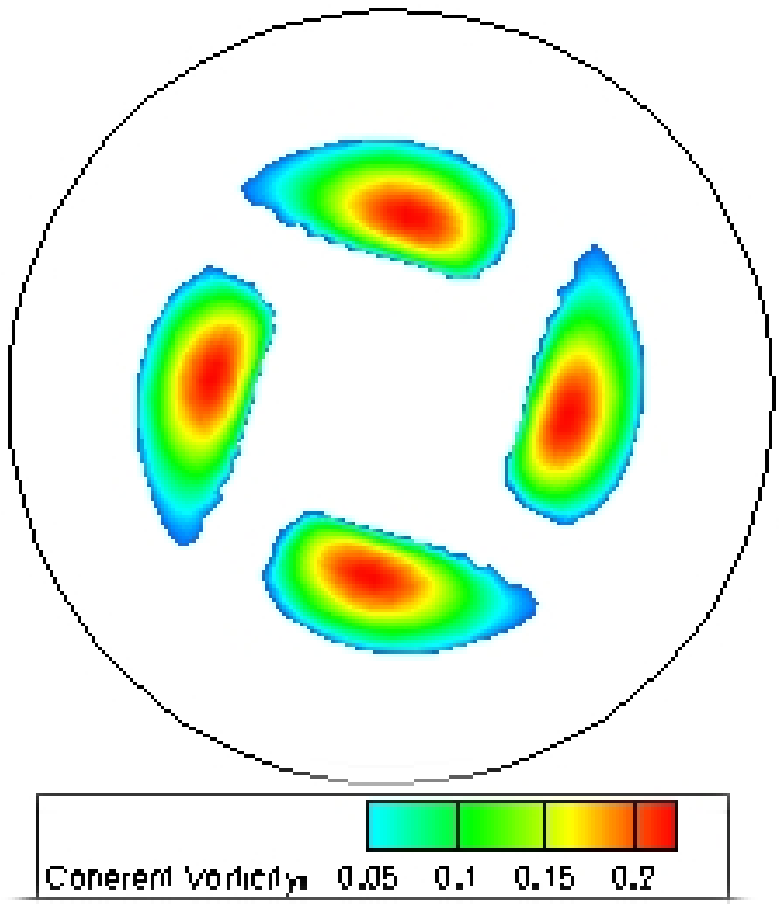}
\end{minipage}
&
\begin{minipage}{1.75 in}
\includegraphics[width=1.75 in]{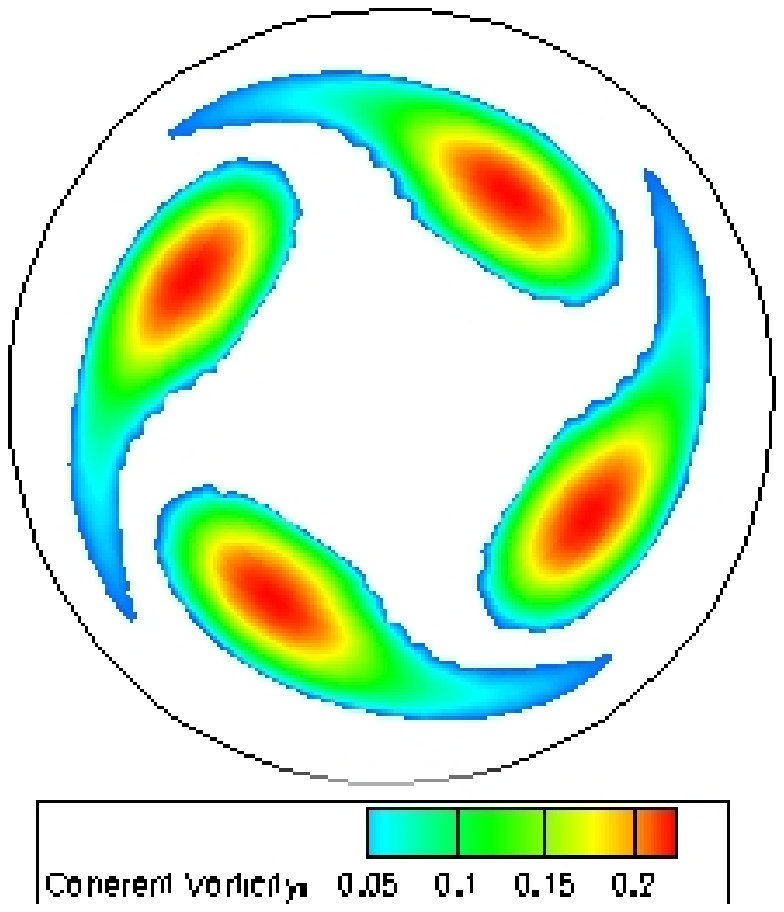}
\end{minipage}\\
\end{tabular}
\caption{The (3,2,1) lift mode with contours of coherent vorticity. Left: (a) $\mathrm{Re}_\tau=95$.
  Right: (b) $\mathrm{Re}_\tau=150$.}
\label{321}
\end{figure}

\begin{figure}[htb]
\begin{tabular}{cc}
\begin{minipage}{1.75 in}
\includegraphics[width=1.75 in]{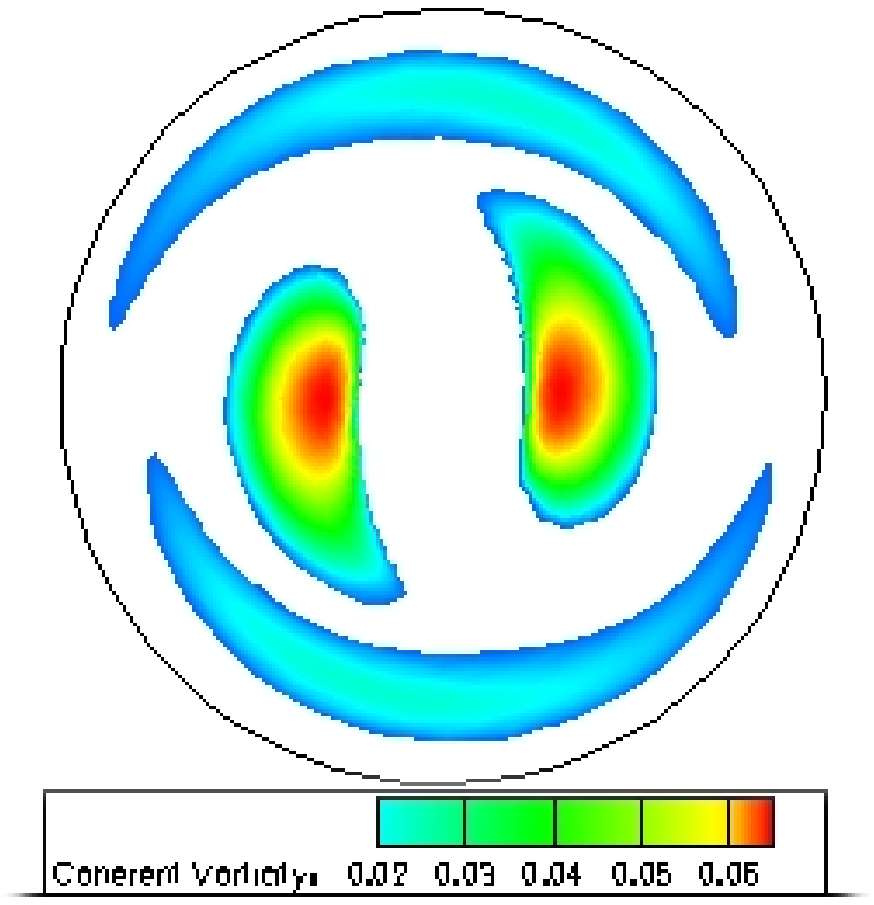}
\end{minipage}
&
\begin{minipage}{1.75 in}
\includegraphics[width=1.75 in]{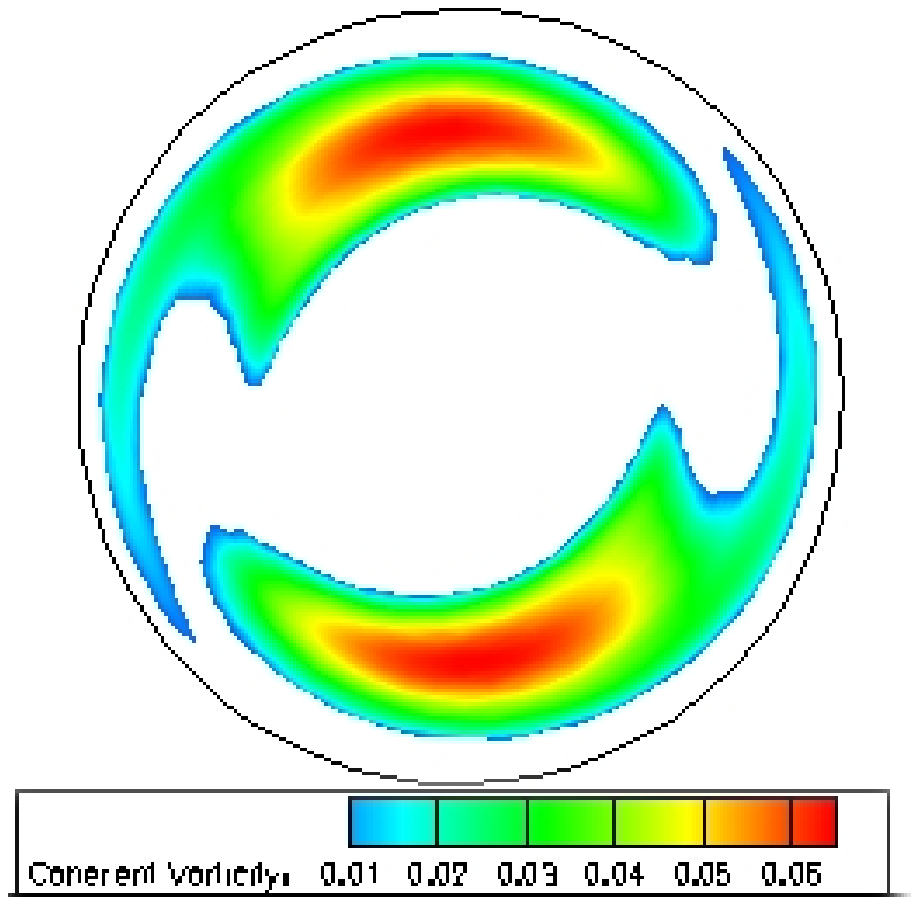}
\end{minipage}\\
\end{tabular}
\caption{The (1,1,1) asymmetric mode with contours of coherent vorticity. Left: (a) $\mathrm{Re}_\tau=95$.
  Right: (b) $\mathrm{Re}_\tau=150$.}
\label{111}
\end{figure}

\begin{figure}[htb]
\includegraphics[width=3.5 in]{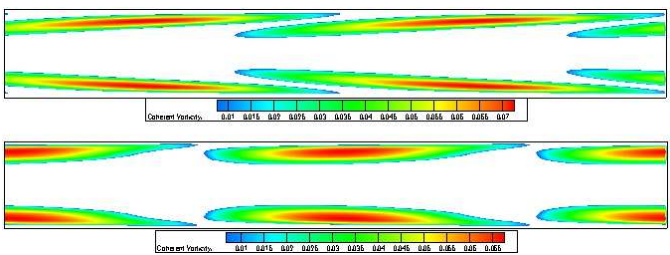}
\caption{The (1,0,1) ring mode ($r$-$z$ plane cross-section) with contours of coherent vorticity. top: (a) $\mathrm{Re}_\tau=95$.
  Bottom: (b) $\mathrm{Re}_\tau=150$.}
\label{101}
\end{figure}

In visualizing the most energetic modes, the same structures as those
found in the $\mathrm{Re}_\tau=150$ case are present.  Figures
\ref{001} - \ref{101} show the coherent vorticity for the most
energetic modes for each of the subclasses discussed in part I.  Since the $(0,0,1)$ mode has no
coherent vorticity, the velocity is shown. Here we use ``coherent
vorticity'' to refer to the imaginary eigenvalues of the velocity
gradient tensor as defined in Chong et al. (also know as ``degree of swirl''). \cite{chong}  Although slight differences
are visible due to the lower Reynolds number, the same trends and
characteristics are found for the KL modes as those found in the $\mathrm{Re}_\tau=150$ case.

\section{DYNAMICS}

\begin{figure}[htb]

\includegraphics[width=2.9 in]{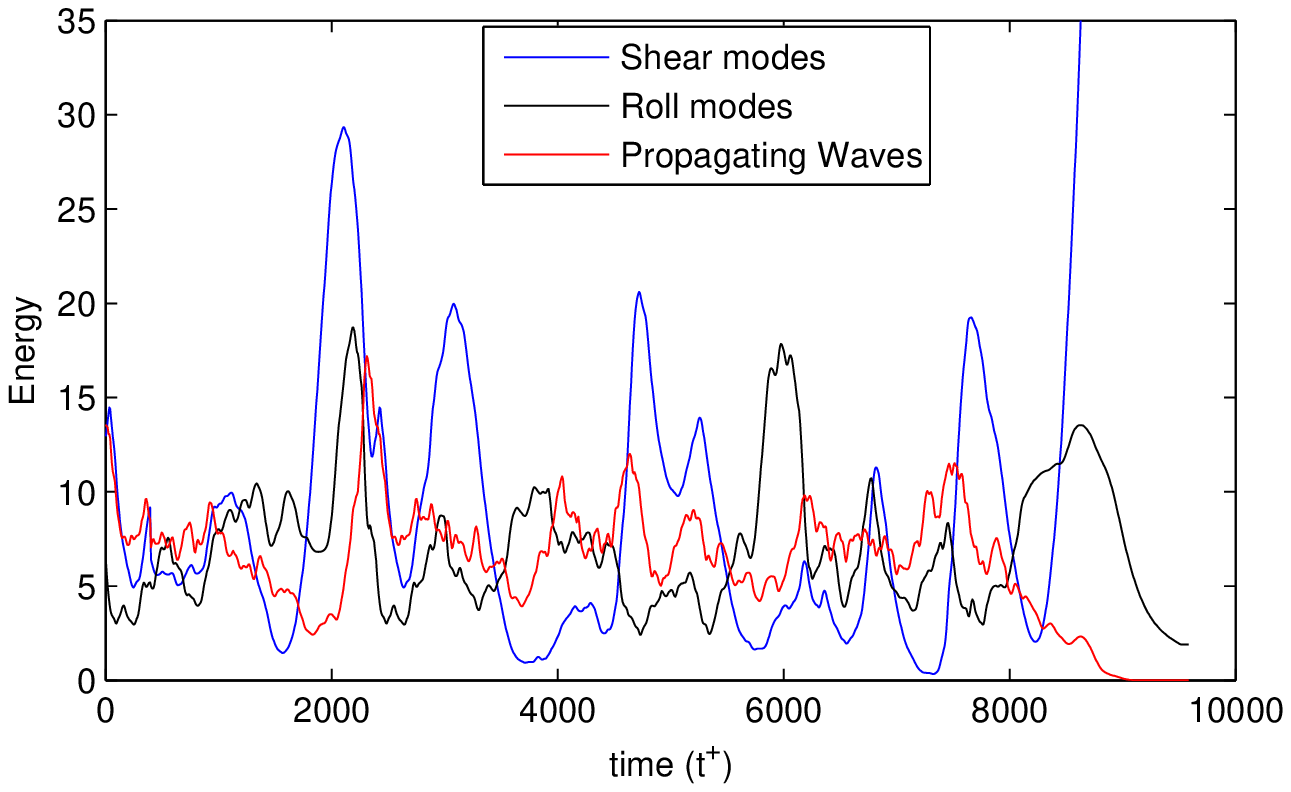}
\caption{Time history of energy of shear modes (blue), roll modes
  (black), and propagating waves (red).}
\label{energyTime}

\begin{minipage}{3.25 in}

\includegraphics[width=2.9 in]{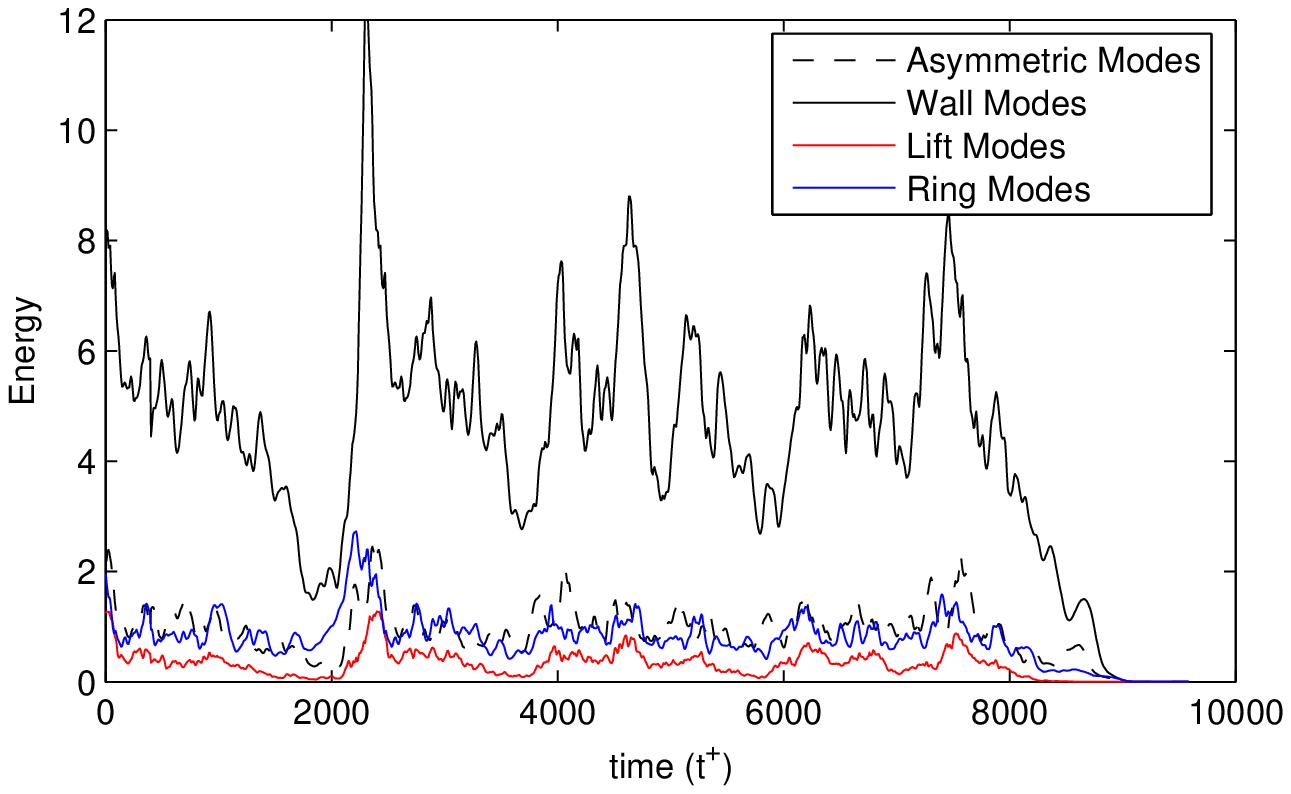}
\caption{Time history of energy of propagating subclasses.  Wall
  (solid black), lift
  (red), asymmetric (dashed), and rings (blue).}
\label{subclassTime}

\includegraphics[width=2.9 in]{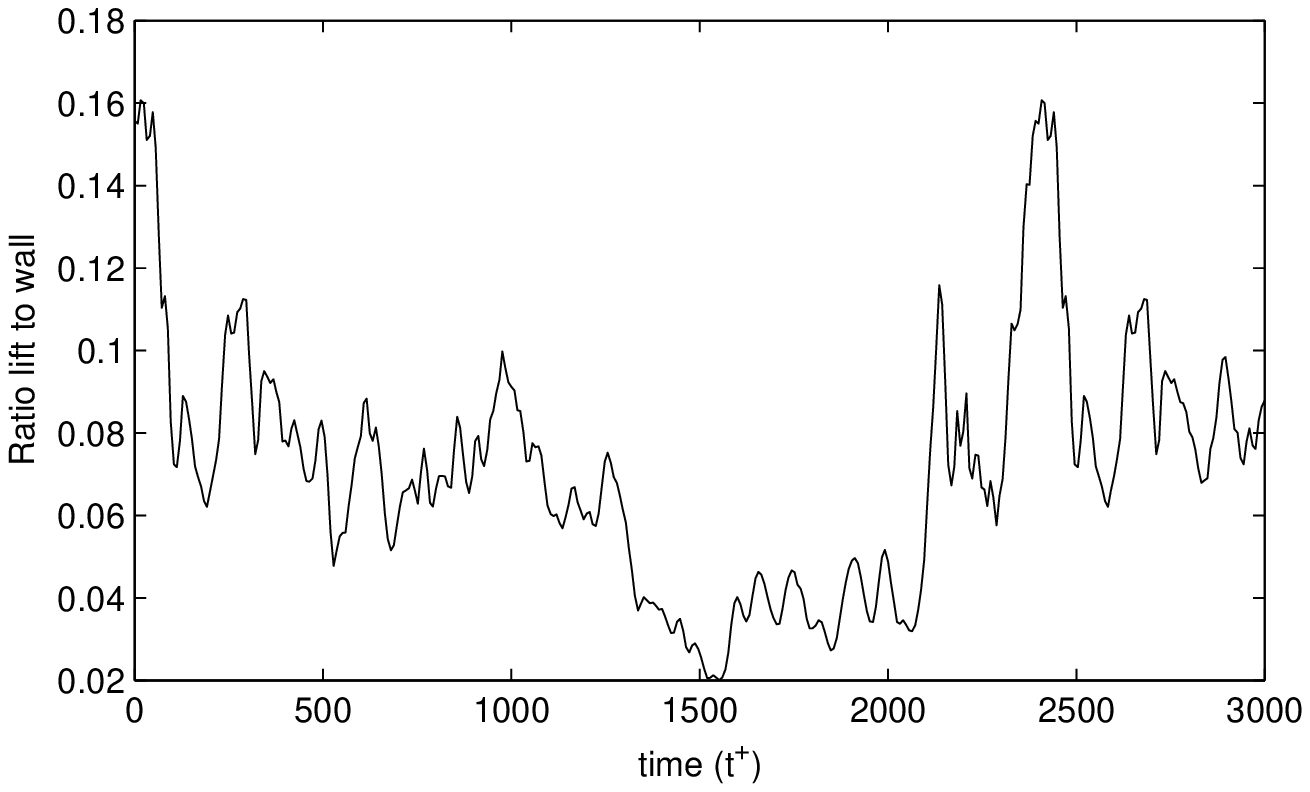}
\caption{Ratio of lift mode energy to wall mode energy as a function
  of time.  As the lift mode energy decreases with respect to the wall
  mode energy, relaminarization begins.}
\label{lift2wall}
\end{minipage}
\end{figure}

Recreating the time history of the KL modes reveals the interaction
between the shear modes, roll modes, and propagating waves.  As shown
in Figure \ref{energyTime}, the chugging phenomena happens when the
propagating modes drop in energy.  This happens at times $t^+ \approx
1600$, $t^+ \approx 3500$, and $t^+ \approx 5500$, and starts the
chugging phenomena.  At $t^+ \approx 8000$, the propagating modes drop
too far in energy, and the flow relaminarizes.  The chugging cycle
ends again when the propagating waves spike at $t^+ \approx 2200$,
$t^+ \approx 4100$, and $t^+ \approx 6200$.  

\begin{figure*}[htb!]
\begin{tabular}{ll}
\begin{minipage}{3.5 in}
\includegraphics[width=3.5 in]{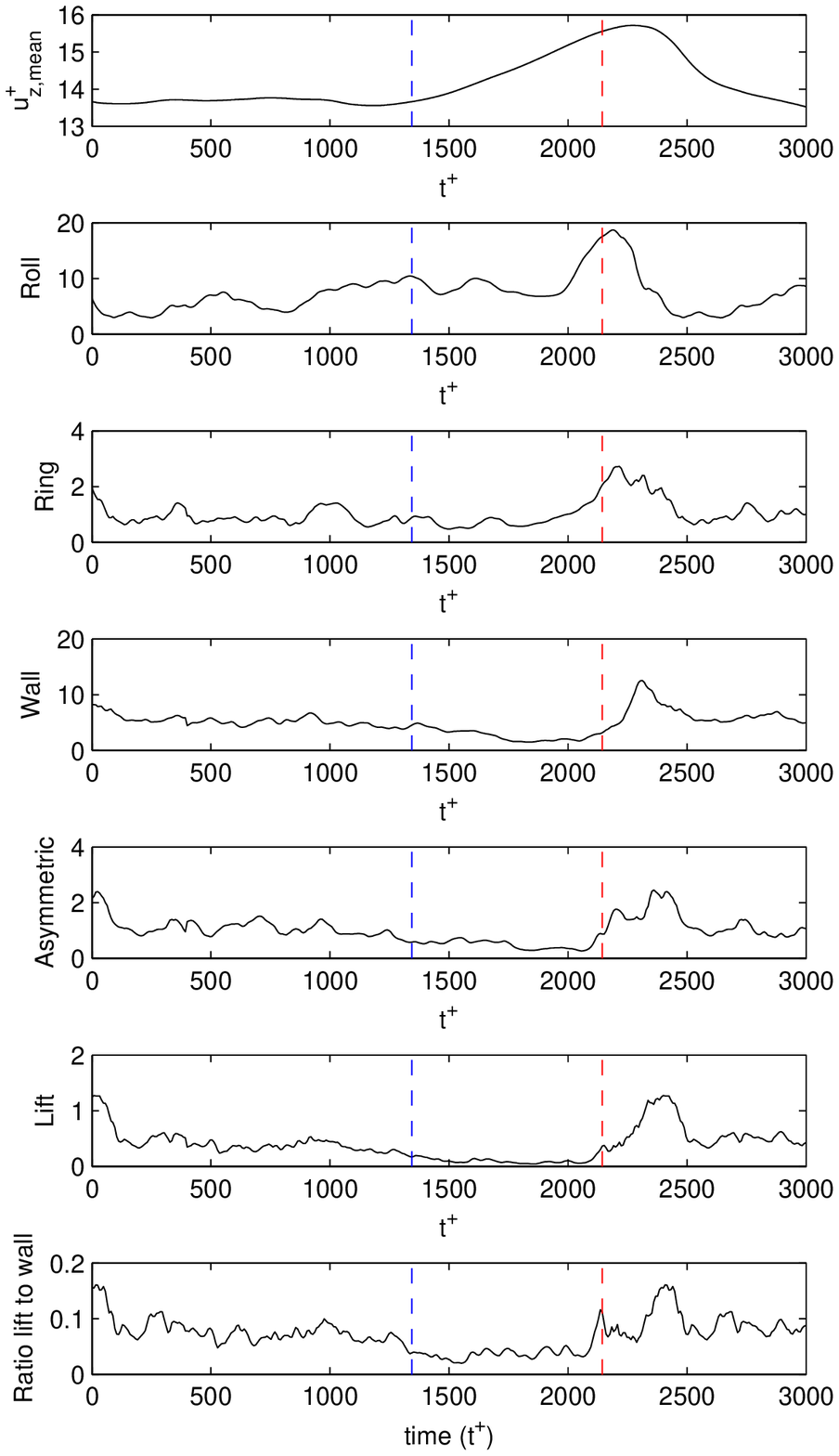}
\caption{Time history of energy of subclasses for a single chug.  From
  the top: the mean velocity, roll mode energy, ring mode energy, wall
  mode energy,
  asymmetric mode energy, lift mode energy, and the lift to wall mode
  energy ratio.  The start of the chug (blue dashed line) is when the lift to wall mode
  energy ratio drops too low.  The end of the chug (red dashed line) is when the ratio
  spikes and recovers.  The phase lag in energy from the roll to the
  wall modes (through the ring modes) and similarly from the wall to
  the lift modes
  (primarily through the asymmetric modes) can be seen. }
\label{oneChugClass}
\end{minipage}
&
\begin{minipage}{3.5 in}

\includegraphics[width=3 in]{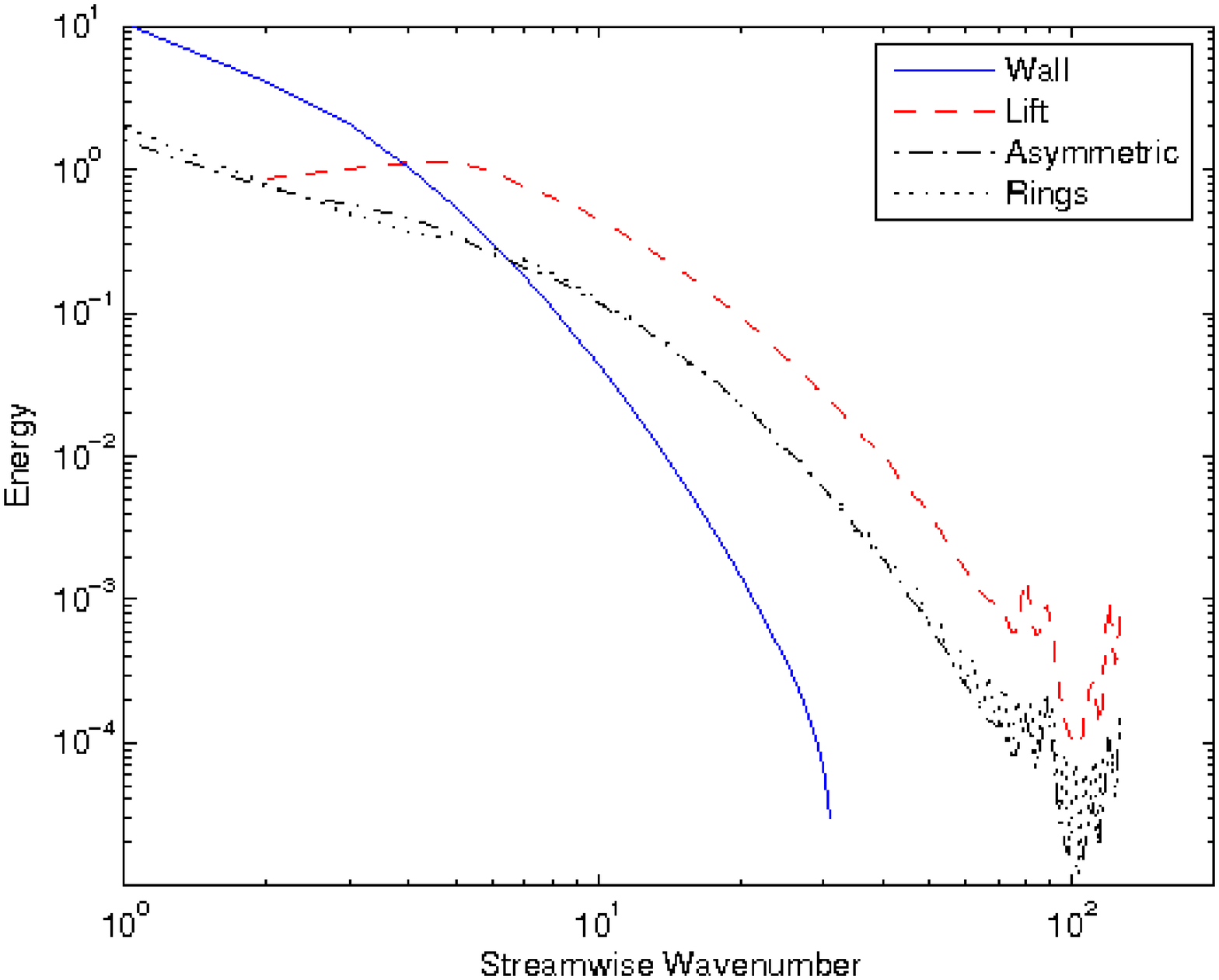}
\caption{Average energy spectra of the propagating modes.  Wall (solid), lift
  (dashed), asymmetric (dots), and rings (dash-dot).}
\label{distribEnergy}

\includegraphics[width=3 in]{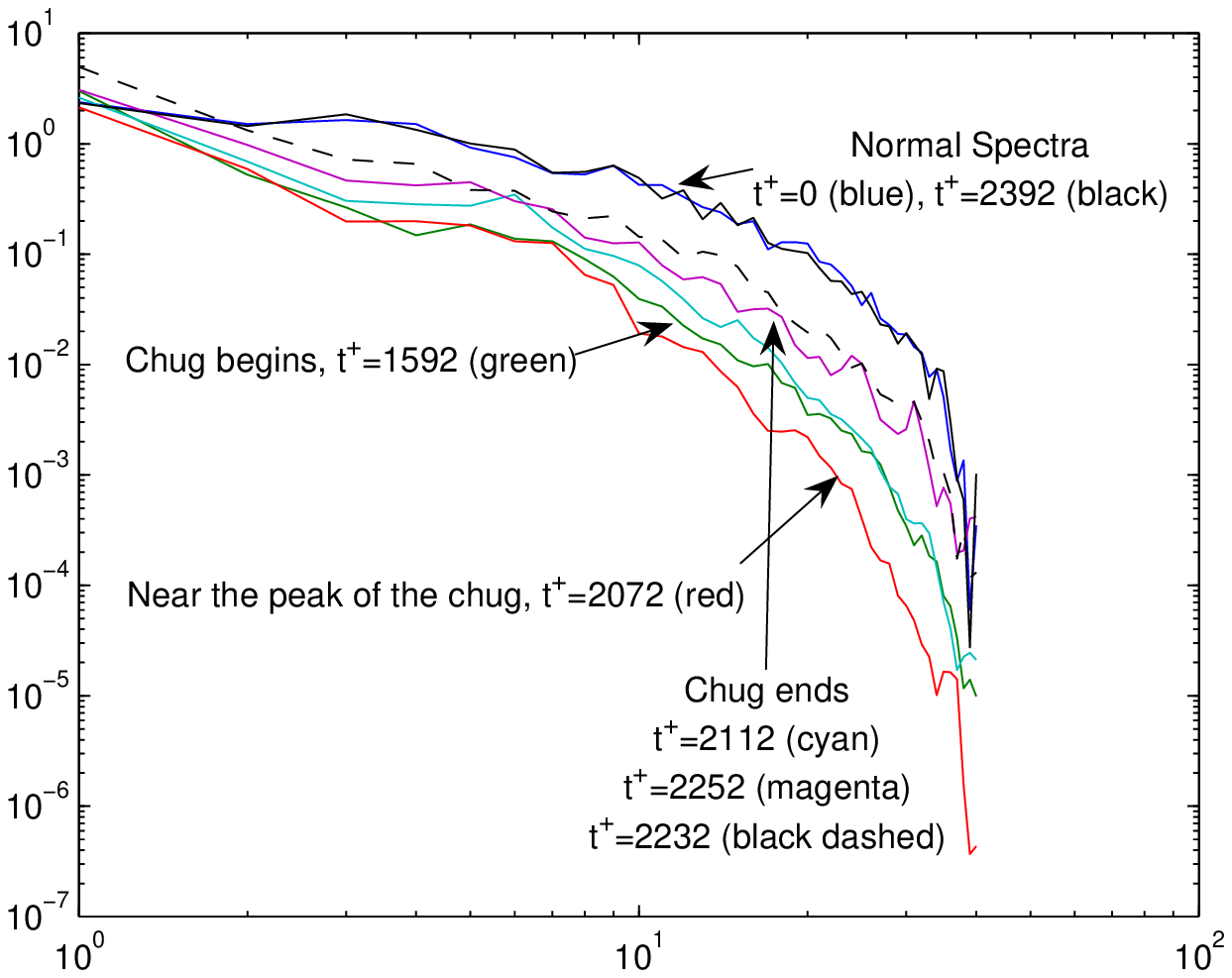}
\caption{Time history of the energy spectra of the propagating modes.  $t^+=0$
  and $t^+=2392$ show the proper established energy spectra at the
  beginning of the simulation, and well after the chug.  $t^+=1592$ is
  the spectra when the chug begins.  $t^+=2112,2552$ and $2232$
  show the spectra regaining strength with time, finishing the
  chug cycle.  Like the dynamics of the lift to wall energy ratio,
  the chugging spectra reinforces the dependence of the self-sustaining mechanism of
  wall turbulence on the high frequencies found in the inertial range,
  represented by the lift modes.}
\label{chugSpectra}
\end{minipage}
\end{tabular}
\end{figure*}

Examining the propagating waves based upon their subclass, shown in
Figure \ref{subclassTime}, we find that the asymmetric and ring modes stay
about the same energy as each other throughout the chugging cycles, while the wall modes are about a
factor of 4 times more energetic.  The lift modes, on the other hand,
vary greatly in energy.  When the lift modes
decay from high to low energy, this coincides with the start of
a chugging cycle, and ends when the lift modes regain the high energy
state.  This is emphasized in Figure \ref{lift2wall} where the ratio
of the lift mode total energy to the wall mode total energy is shown.
In Figure \ref{oneChugClass}, only one chug is shown, emphasizing the
importance of the lift to wall ratio, and also the energy transfer
between classes as energy flows from the rolls to the wall modes
(through the ring modes), and then from the wall modes

\newpage

\noindent  to the lift modes (primarily
through the asymmetric modes).

Observing the energy spectra throughout the chug, we first revisit the total
energy distribution of the propagating waves as averaged over the
entire flow, seen in Figure \ref{distribEnergy}.  Again, like the
$\mathrm{Re}_\tau=150$ spectrum, the lift modes are more energetic
than the wall modes for high wavenumbers.  Thus, because
the high wavenumbers of the spectra dies off by two orders of magnitude at the start of the chug cycle,
seen in Figure \ref{chugSpectra}, it reinforces the importance of the
lift modes in maintaining the turbulent flow.

\begin{figure}[h]
\fbox{
\includegraphics[width=3 in]{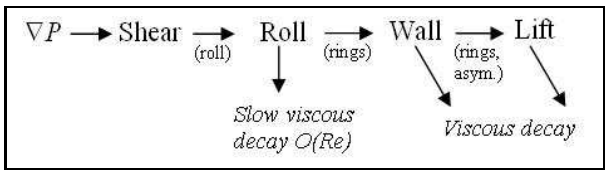}}
\caption{Energy flow chart for turbulence subclasses with catalysts.
  If any of these legs are disrupted, turbulence reduction (or drag
  reduction) begins. For instance, in Part I
  the energy in the wall modes 
  were reduced by lifting them off the wall by spanwise wall
  oscillations, reducing their energy by forcing them to
  advect faster and died faster.  In the current relaminarization case, not
  enough energy is present to fully populate the lift modes, and so their
  energy decreases and again the energy balance is disrupted.}
\label{energyDiagram}

\end{figure}

As noted in the $\mathrm{Re}_\tau=150$ case, the lift modes are
responsible for the majority of turbulence near the center of the pipe,
as the wall modes stay near the wall, even for high quantum number.
Thus, the importance of the outer region in the self
sustaining mechanism of turbulence is reinforced.  If the lift modes
do not receive enough energy, cascaded through the wall modes, the
relaminarization process begins near the center of the pipe.  This is confirmed
by the presence of the large drop in the rms velocities and Reynolds stress
profiles near the center of the pipe.  If the
process cannot be halted in 
time by the transfer of energy from the wall modes to the lift
modes, typically through the (1,3,1) or (1,4,1) modes, the
flow will completely relaminarize.  Adding in the findings of triad
interactions  by Webber et al., \cite{webber2} the flow of energy is
shown in Figure \ref{energyDiagram}.  The
shear to roll interactions are catalyzed by the 
rolls themselves, the roll to wall interactions are catalyzed mostly by the
ring modes, and the wall to lift modes are catalyzed by both the ring and
the asymmetric modes, and other lift modes.  In the relaminarization
process, it is this final leg that fails, breaking the mechanism, and
starting relaminarization from the center of the pipe.  

\section{CONCLUSIONS}

It is apparent through the examination of the spanwise wall oscillated
case in Part I and the relaminarization case in Part II that if any leg of the energy cycle
in a turbulent flow is disrupted, the resulting imbalance can lead to
the start of a relaminarization process, and even complete
relaminarization.  In Part I, we describe a model where the propagating wall modes
were pushed to higher advection speeds, reducing their effective
lifespan.  Thus they do not have enough time to take energy from the roll modes, breaking the third leg of the mechanism.  Here in Part II, with
lower pressure gradient, there is not 
enough energy to properly maintain the lift modes, and the last leg of
the process is broken, starting the
relaminarization process. 

 Thus, in conclusion, we find that while the wall modes and near wall
 interactions are responsible for the generation of turbulence from
 the pressure gradient, the turbulence in the
outer region is necessary to maintain the proper inertial
range in the energy spectra, and that without it, the relaminarization
 process begins.

\section*{ACKNOWLEDGMENTS}
This research was supported in part by the National Science Foundation
through TeraGrid resources provided by the San Diego Supercomputing
Center, and by Virginia Tech through their Terascale Computing
Facility, System X.   We gratefully acknowledge many useful interactions with Paul Fischer
and for the use of his
spectral element algorithm.

\bibliography{references}
\end{document}